# Deep Keck adaptive optics searches for extrasolar planets in the dust of Epsilon Eridani and Vega

Bruce A. Macintosh

*Institute of Geophysics and Planetary Physics, Lawrence Livermore National Laboratory, 7000 East Ave. L-413, Livermore, CA 94551*

`bmac@igpp.lllnl.gov`

E. E. Becklin, Denise Kaisler, Quinn Konopacky, B. Zuckerman

*Department of Physics and Astronomy, UCLA, 8371 Math Sciences Building, Box 951562 Los Angeles, CA 90095-1562*

## ABSTRACT

A significant population of nearby stars have strong far-infrared excesses, now known to be due to circumstellar dust in regions analogous to the Kuiper Belt of our solar system, though orders of magnitude more dense. Recent sub-mm and mm imaging of these systems resolves the circumstellar dust and reveals complex structures, often in the form of rings with azimuthal non-axisymmetric variations. This structure might well be due to the presence of embedded brown dwarfs or planets. We have carried out deep adaptive optics imaging of two nearby stars with such asymmetric dust: Epsilon Eridani and Vega. Ten and seven candidate companions were seen in and near the dust rings of Epsilon Eridani and Vega respectively, but second-epoch proper motion measurements indicate that all are background objects. Around these two stars we can thus exclude planetary companions at spatial scales comparable to the radius of the dust structures to a level of $M_K$=24, corresponding to 5 Jupiter masses, for Epsilon Eridani, and $M_K$=19-21, corresponding to 6-8 Jupiter masses, for Vega.

## 1. Introduction

The IRAS satellite discovered that a significant population of nearby main sequence stars, including Vega, display strong excess far-infrared emission, now known to be due to circumstellar dust (Zuckerman 2001 and references therein.) The region containing the dust at these "Vega-like" stars is analogous to the Solar System region associated with the Kuiper Belt, though the total dust mass is orders of magnitude higher than in our system. Recent sub-mm and mm imaging of these systems (Holland et al. 1998, Greaves et al. 1998, Koerner et al. 2002, Willner et al. 2002, Holland et al. 2002) resolves the circumstellar dust and reveals complex structures, often in the form of rings with azimuthal variations. This structure might be due to the presence of embedded brown dwarfs or planets (Liou & Zook 1999; Ozernoy et al. 2000, Kuchner and Holman 2002). Using the 10-m W.M. Keck II telescope we have carried out deep near-infrared adaptive optics (AO) imaging of the regions near two stars with such asymmetric dust, Epsilon Eridani and Vega, to search for planetary companions responsible for the structure in the dust ring.

## 2. Properties of the target stars

The properties of the target stars are summarized in Table 1.

### 2.1 Epsilon Eridani

Epsilon Eridani is a particularly fascinating system from the standpoint of extrasolar planet detection. At 3.3 pc, it is one of the closest stars to Earth. Eps Eri is thought to have a relatively young age of ~730 Myr (Song et al. 2000), and has a mass similar to that of our sun. There is also tentative radial-velocity evidence for a companion in an a=3.4 AU orbit (Hatzes et al. 2000). These factors all combine to make it an attractive target for a sustained, deep imaging search for an extrasolar planet.

Epsilon Eridani's circumstellar dust was first resolved by Greaves et al. 1998). As the closest "dusty star", even the low resolution of SCUBA shows a well-defined ring structure with a radius of 20-30 arcseconds and apparently several dense regions. The presence of this considerable substructure suggests the possibility of an unseen agent, most likely a low-mass companion, shaping the dust. Simulations of the resonant structures in such a ring (e.g. Ozernoy et al. 2000 and Liou & Zook 1999) indicate that, to produce such structures, a companion would have to be located not in the midst of the dust "lumps" but behind or ahead of them. COBE observations of zodiacal dust in our solar system show similar structures caused by the Earth (Reach et al. 1995). The modeling work does not constitute a unique solution, of course, so the best strategy is to completely image the circumstellar environment over a range of separations. Epsilon Eridani has a high proper motion ($pm_{ra}$ ~1"/year), so a single epoch of follow-up observations has allowed us to distinguish any true companion from a background object.

**2.2 Vega**

Vega, 8 pc from Earth, is the archetypal early-type infrared-excess star (Aumann et al. 1984). Sub-mm imaging (Holland et al. 1998) does not show a well-defined ring but instead two concentrations of emission at 10-15 arcseconds. These could be either strong inhomegeneities in a face-on dust structure or the ends of a nearly-edge-on ring; Vega itself is thought to be pole-on (Gulliver et al. 1994.) Millimeter interferometry (Koerner et al. 2002, Wilner et al. 2002) confirms the existence of these bright areas, with spectral properties consistent with warm dust. Modeling (Gorkavyi and Taidakova 2001 and in Wilner et al. 2002) shows that, as with Epsilon Eridani, this structure could be produced by a single planetary companion.

**3. Observations**

We observed both target stars with the NIRC2 camera and the facility AO system (Wizinowich et al. 2000a, 2000b) on the W.M. Keck II telescope. NIRC2 has a 1024x1024 pixel array and two main plate scales – 0.01"/pixel and 0.04"/pixel. Although the 0.04"/pixel scale marginally undersamples the typical K'-band image FWHM (0.06"), it is still a better choice for maximizing the (sensitivity•area) product than the oversampled 0.01"/pixel mode. This provides a field of view of 40" per image, corresponding to 130 AU at Epsilon Eridani and 320 AU at Vega.

Epsilon Eridani was first observed on December 1 and 21, 2001 (UT). We observed four fields, offset 24" N, S, E, and W from the star itself; this placed Epsilon Eridani off the array, minimizing ghost images and internally scattered light. The AO system remained locked on the

V=3.7 star, producing good AO correction (Strehl ~ 0.4 at 2.1 μm). Although NIRC2 has a focal-plane coronagraph and selectable Lyot pupil stops, these modes had not been fully commissioned and we did not use them for these observations.

At each position we obtained 15 K' (2.1 μm) images in a 5-position dither pattern, each consisting of 6 coadds of 15 seconds exposure, for a total of 22.5 minutes of integration. Since the edge of each image was dominated by bright scattered light from Epsilon Eridani itself, we had to obtain separate sky images, 10-15 images per target set, in positions offset by 600" from the star. Observations are summarized in Table 2, and Figure 1 shows the locations of the fields observed. Conditions were excellent and photometric for the December 1 observations, but somewhat non-photometric (estimated at ~1 magnitude extinction) during the December 21 observations covering the northern field.

Vega was observed in February 2002 (see Table 2.) Since the angular extent of the dust structures near Vega is smaller than the dust extent at Epsilon Eridani and the Vega dust is asymmetric, we observed only two positions: one centered on Vega, with the star placed behind a 2" diameter partially-transparent occulting spot, and one offset 5" N and E, covering the regions where the dust is densest. The pointing centered on Vega probes a similar physical scale (<160 AU) to the four Epsilon Eridani images, and the offset image provides additional phase space in the direction where the dust is denser, though of course a perturbing planet need not be located inside the dust itself. The observations were otherwise identical to those taken of Epsilon Eridani, with 15 x 6 x 15 seconds of total exposure per position. Sensitivity was similar in both the centered image and the offset image; the occulting spot by itself provided no significant rejection of scattered light. This is unsurprising since the main source of the scattered light halo at these large radii is residual atmospheric or telescope phase errors uncorrected by the adaptive optics system rather than diffraction.

The dust rings themselves are, unsurprisingly, invisible in these near-infrared images. The total mid-IR optical depth in dust near Vega is more than an order of magnitude lower than that of dust disks that have been detected in scattered light, such as Beta Pictoris. In addition, AO observations are ill-suited to circumstellar dust detection. Adaptive optics only reduces light scattered by the atmosphere at separations < ~$\lambda/d$ (where d is the subaperture size of the AO system), which is ~0.7 arcseconds for the Keck AO system. Beyond this radius the scattered light halo is essentially the same as in a non-AO observation. Although AO can still provide enormous gains in point-source sensitivity by concentrating the light of a possible companion into a diffraction-limited spike, it provides insignificant enhancement to sensitivity to diffuse circumstellar emission. Thus, although our sensitivity to point sources is considerably greater than the NICMOS observations of Silverstone et al. (2002), our sensitivity to diffuse emission is actually less.

Ten point sources were detected near Epsilon Eridani and seven near Vega. Both stars were re-observed in August 2002. Total exposure times were the same; data obtained on August 20 were near-photometric, data from August 21 of somewhat lower quality. Due to poor conditions we were unable to re-observe the northern Epsilon Eridani offset field. Only one source in this field is not in the region of overlap with the eastern and western fields, and that source is near the northern edge of the northern field and hence highly unlikely to be a companion. All other sources detected in the first-epoch images were re-detected in the second. The second-epoch observations of Vega are offset further north and east than the first-epoch and detect a new source near the eastern edge, but this is again unlikely to be a true companion.

## 4. Data analysis and results

Images were dark-subtracted, sky-subtracted and flat-fielded with standard infrared astronomical techniques. Dithered images in each field were registered by measuring the positions of point sources present in them, and median-combined to reject artifacts and ghosts. The images were then processed with an "unsharp mask", by subtracting a median-smoothed version of the image from itself; this has the effect of removing any smooth scattered light background and highlighting point sources.

We then identified and measured the positions of all the apparently-pointlike sources in each field. All of the objects seen in the field appear to be point sources within the resolution of the AO system (~0.1", including undersampling and isoplanatism effects.)

Figures 2 and 3 show the images with the point sources numbered. We measured approximate offsets from the (highly saturated) image of the primary star or the point of intersection of the diffraction spikes, but for astrometric purposes we measured the offsets of the point sources relative to the brightest source in each field, rather than to the primary, which was typically off the field or saturated. In the second-epoch images, each target was re-identified and its position remeasured. Table 3 summarizes the positions of the identified point sources and their change in position since the first epoch. Distortions in the camera were corrected using the equations given in the NIRC2 pre-ship review (http://alamoana.keck.hawaii.edu/inst/nirc2/preship/preship_testing.ps); these were not significant for Epsilon Eridani, in which each field was observed in the same orientation in each epoch, but were significant for Vega, in which Keck image rotator was oriented differently during the second epoch. Based on measurements of relative positions of the brightest sources in multiple images the expected uncertainty in the relative astrometry is estimated to be ±0.03 arcseconds.

Since it is highly unlikely that all objects would be planetary companions, by using the brightest object in each field as a reference grid we would expect to measure a change in position equal to the proper motion of the primary star; in fact, within the errors, no candidates changed their relative positions. Vega source 7 changed position at the 3 sigma level, but not in the direction expected for a true companion. This source is located near the edge of the image and may be subject to residual distortion effects.

We performed coarse aperture photometry of our targets. Aperture photometry on images of the star HD77281 were used for absolute calibration, with an 0.8" radius aperture, large enough that variations in AO performance won't affect the photometric zeropoint (although variations in seeing could still change the calibration.) HD77281 is sufficiently bright (V=7) that AO performance was comparable to that on Epsilon Eridani or Vega. We then used the same aperture to measure the brightness of source 2 in the Epsilon Eridani images. Relative photometry for all the sources was determined using the unsharp-mask images (to remove the effects of diffuse background light from the primary star) and 0.12" radius apertures, with the calibration tied to the measurement of source 2. Since the differences between the point spread function of the AO system during Epsilon Eridani and photometric standard observations are unknown, and since isoplanatic effects may further reduce the Strehl ratio at large radii, the relative accuracy between different companions (especially those at similar radii) should be good but with ±0.3 magnitudes of error in absolute calibration. For Vega, we followed a similar procedure using source number 1.

Our data can be used to set upper limits on planetary companions near these stars. We measured the noise in an image at a given radius by calculating the standard deviation in narrow annuli, scaled to the size of our photometric apertures and compared to an estimate of the flux in the core of the AO PSF from the photometric calibration discussed above. The resulting 5-sigma limiting magnitude is shown in Figure 3. Except for the large diffraction spikes (which are partially removed through combination of multiple images) our images are relatively uniform – sensitivity does not vary as a function of azimuth – though sensitivity is significantly lower in the northern offset field of Epsilon Eridani due to poor conditions. The Vega sensitivity limits come from the offset image, which has comparable sensitivity as a function of radius to the centered image. These can be compared to the predicted brightness of extrasolar planets from models (Burrows et al. 1997, Burrows 2002, Marley 2002.) We could detect a 4-5 Jupiter-mass planet at the separation of the Epsilon Eridani dust ring and a 6-8 Jupiter-mass planet at the separations of the Vega dust structure.

It is interesting to compare our results to those of Metchev et al. (2002), who observed Vega with the Palomar AO system (PALAO). Our sensitivity is 2-3 magnitudes greater, but in terms of detectable companion mass is roughly comparable to what they claim. This is largely because they have observed at H band; brown dwarf models predict extremely blue H-K colors, such as -2.1 for a $T_{eff}$=450 K 6 Jupiter-mass object at the age of Vega (Burrows et al. 1997). However, factors such as clouds (Marley et al. 2002) can operate to bring objects closer to black-body spectra; for observed brown dwarfs, clouds do not seem to be significant below $T_{eff}$=1200, but for lower-gravity objects such as planets their strength is unknown. If clouds are significant, the H-K colors of planets would be redder and hence our mass limits would be lower than those of Metchev et al. Since the properties of planetary-mass objects in this temperature range are unknown, observing at a range of different wavelengths may be a sensible strategy and the data of our two groups complement each other.

Over the overlap between our fields and those of Metchev et al., we detect all sources in their images; our source 4 is in their field but below their sensitivity limit. Our photometry, though crude, systematically disagrees with theirs by approximately 0.5 magnitudes. Absolute adaptive optics photometry is notoriously difficult, so this could be due to differences in AO performance between their photometric calibration and Vega observations. Their seeing was described as mediocre and variable (0.7-1.0" in H) and they used a V=10 calibrator, which might cause significantly worse AO performance for an AO system with small subapertures such as PALAO. McCarthy (2001) used conventional near-IR imaging to search for substellar companions to young stars, and also observed Vega in conjunction with Holland et al. (1998). Although direct imaging is much less sensitive than AO imaging it is also photometrically easier to calibrate, and our photometry for the brightest sources near both Vega and Epsilon Eridani is consistent with that of McCarthy (2002 priv. com.)

## 5. Conclusions:

Deep AO imaging of fields around Epsilon Eridani and Vega show no evidence of brown dwarf or planetary companions that could be confining or shaping the dust ring, down to the 5 Jupiter-mass level (Epsilon Eridani) and 6-8 Jupiter-mass level (Vega) at the angular separations comparable to that of the dust rings. It is worth noting that our sensitivity was continuing to increase as $t^{1/2}$ during our observations – ie, no systematic effects were limiting sensitivity at these large separations – and hence deeper imaging in the future could reach the 2-3 Jupiter mass level that some authors (Kuchner and Holman 2003) have predicted for the planet near Vega.

**Acknowledgements:**

The authors would like to thank David le Mignant for assistance with the Keck adaptive optics system during these observations, and Chris McCarthy for providing near-IR images of Vega and Epsilon Eridani for comparison to our data. We would also like to thank the anonymous referee for a number of helpful comments. This research was performed under the auspices of the U.S. Department of Energy by the University of California, Lawrence Livermore National Laboratory under Contract W-7405-ENG-48, and also supported in part by the National Science Foundation Science and Technology Center for Adaptive Optics, managed by the University of California at Santa Cruz under cooperative agreement No. AST – 9876783. This research was also supported in part by the UCLA Astrobiology Institute and by a NASA grant to UCLA. Data presented herein were obtained at the W.M. Keck Observatory, which is operated as a scientific partnership among the California Institute of Technology, the University of California and the National Aeronautics and Space Administration. The Observatory was made possible by the generous financial support of the W.M. Keck Foundation.

Table 1: Properties of target stars

|  | Epsilon Eridani | Vega |
|---|---|---|
| Spectral type | K2V [1] | A0V [1] |
| Distance (pc) | 3.22 [1] | 7.76 [1] |
| V magnitude | 3.73 [1] | 0.03 [1] |
| K magnitude | 1.62 [1] | -0.06 [1] |
| Age (MYr) | 730 [2] | 350 [3] |
| Proper motion (RA, DEC) arcsec/year | -0.98, 0.02 [1] | 0.20, 0.29 [1] |

1. From the SIMBAD database

2. Song et al. 2000

3. Barrado y Navascues 1998, Song et al. 2001

Table 2: UT Dates of observations

| Star | Field offset | First Epoch | Second Epoch |
|---|---|---|---|
| Epsilon Eridani | 24" E | 01 Dec. 01 | 20 Aug. 02 |
| Epsilon Eridani | 24" S | 01 Dec. 01 | 20 Aug. 02 |
| Epsilon Eridani | 24" W | 01 Dec. 01 | 21 Aug. 02 |
| Epsilon Eridani | 24" N | 21 Dec. 01 | |
| Vega | center | 21 Feb. 02 | |
| Vega | 5" N 5" E | 21 Feb. 02 | |
| Vega | 10" 10" E | | 20 Aug. 02 |

Table 3a: Candidate companions near Epsilon Eridani

| Object | Offset RA from primary (arcseconds) | Offset Dec from primary (arcseconds) | Astrometric reference | $\Delta$ RA | $\Delta$ DEC | $m_{K'}$ |
|---|---|---|---|---|---|---|
| Expected motion of a true companion | | | | -0.704 | 0.014 | |
| 1 | -9.6 | 14.2 | ref for N, W | ref | ref | 17.3 |
| 2 | 4.5 | 17.0 | ref for E | ref | ref | 17.3 |
| 3 | -4.2 | 44.1 | 1 | * | * | 16.3 |
| 4 | 26.8 | 10.3 | 2 | 0.018 | 0.009 | 19.4 |
| 5 | 38.7 | 14.5 | 2 | 0.096 | 0.018 | 20.7 |
| 6 | 36.1 | -19.6 | 2 | 0.032 | 0.018 | 20.2 |
| 7 | -13.1 | 16.4 | 1 | 0.016 | 0.056 | 20.3 |
| 8 | -31.2 | 13.6 | 1 | 0.001 | 0.026 | 20.1 |
| 9 | -17.8 | -34.1 | ref for S | ref | ref | 19.3 |
| 10 | 15.9 | -15.9 | 9 | 0.020 | 0.033 | 20.8 |

Table 3b: Candidate companions near Vega

| Object | Offset RA from primary | Offset Dec from primary | Astrometric reference | Δ RA | Δ DEC | $m_K$ |
|---|---|---|---|---|---|---|
| Expected motion of a true companion | | | | 0.098 | 0.143 | |
| 1 | 21.6 | -5.0 | Ref for all | --- | --- | 14.9 |
| 2 | 20.2 | -0.5 | 1 | -0.027 | 0.020 | 17.2 |
| 3 | 27.5 | -2.9 | n/a | ** | ** | 18.5 |
| 4 | 25.5 | 8.8 | 1 | -0.005 | 0.028 | 19.4 |
| 5 | 22.6 | 18.8 | 1 | 0.012 | 0.038 | 16.3 |
| 6 | 20.1 | 18.9 | 1 | 0.023 | 0.052 | 20.5 |
| 7 | 11.9 | 24.6 | 1 | 0.110 | 0.008 | 18.3 |

*: No observations in second epoch

**: No observations in first epoch

Columns: Offset RA and Dec are offsets from primary star in the first measurement epoch. Typical errors are +-0.2 arcseconds, dominated by uncertainty in the position of the primary star. Astrometric reference indicates either which field the star was used as a reference for or which candidate was used for the measurements of the change in position between the two epochs. Δ RA and Δ DEC indicate the change in relative position of the candidate between the two epochs. Uncertainties in Δ RA and Δ DEC are 0.02 arcseconds. $m_K$ gives an approximate apparent K magnitude; errors (mainly due to uncertainties in the quality of AO correction and in isoplanatic effects) are ± 0.3.

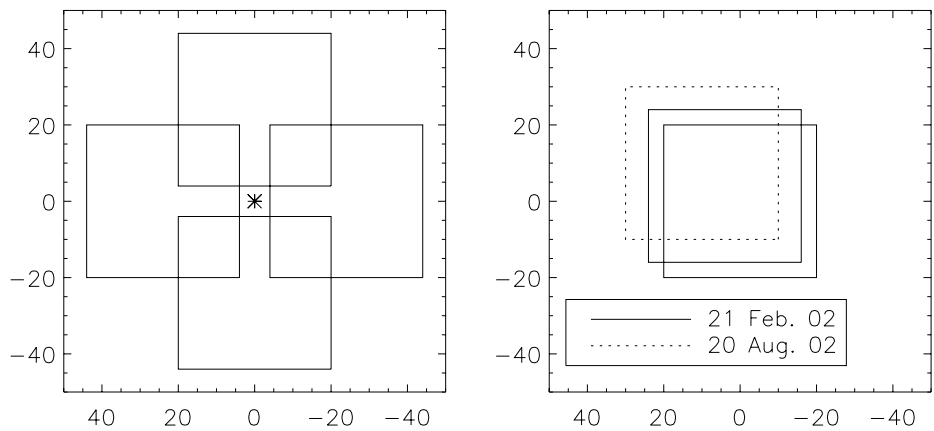

Figure 1: Field of view of the four Epsilon Eridani images (left) and two Vega images (right). Axes are in arcseconds.

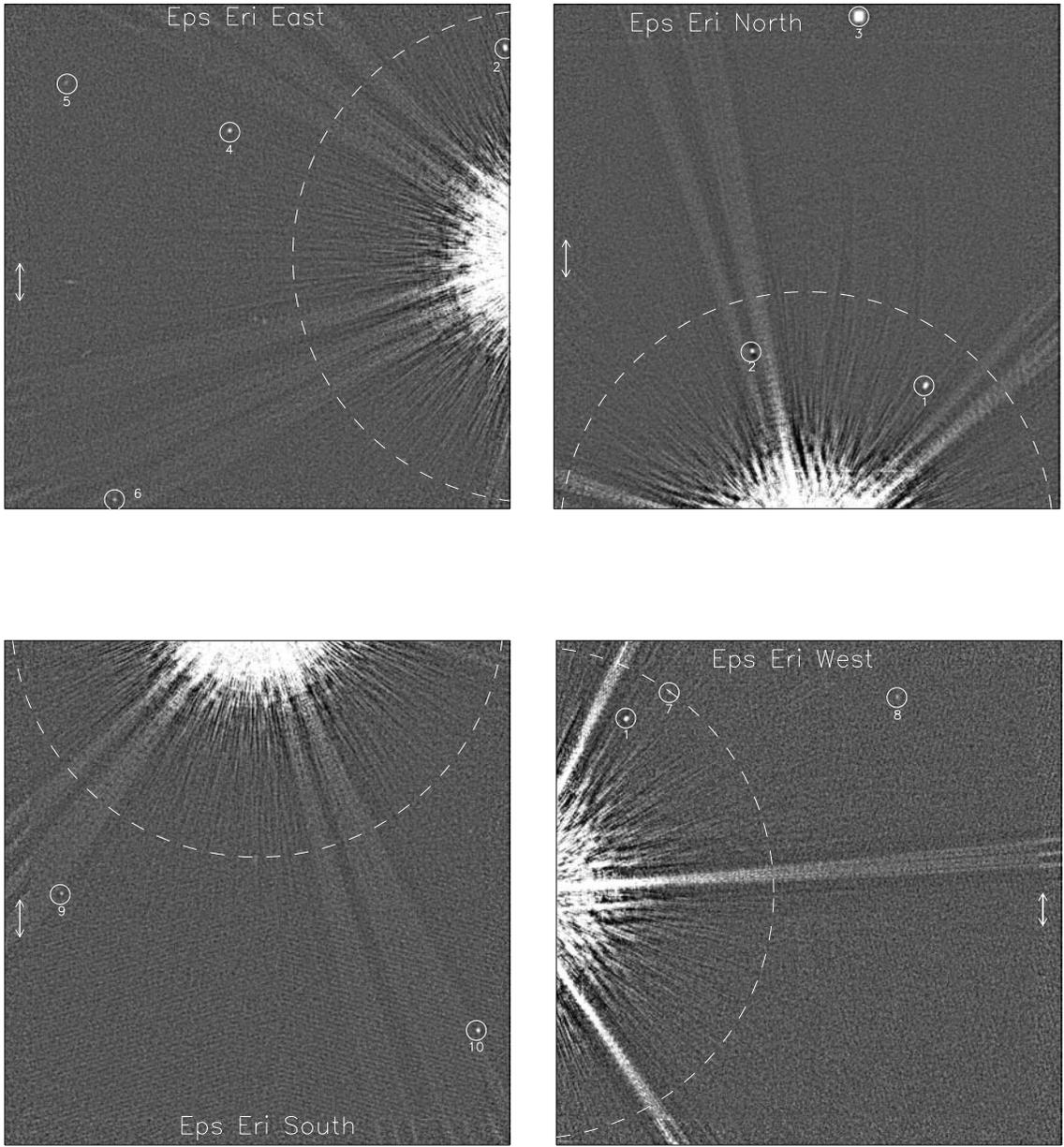

Figure 2a-d: Deep NIRC2 K' images of Epsilon Eridani, offset 24" E (a, upper left), N (b, upper right), S (c, lower left) and W (d, lower right) from the star. The dashed line indicates a radius of 20 arcseconds from the primary star, and the arrow a length of 3". Candidate companions (all now known to be background objects) have been circled.

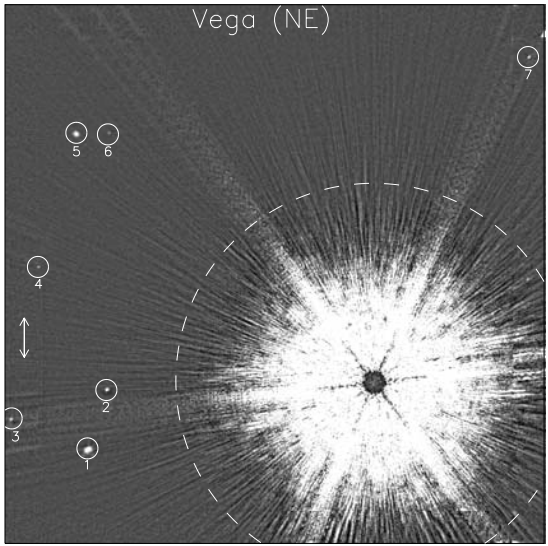

Figure 3: Deep NIRC2 image of the second-epoch field around Vega, offset 10" N and E of the primary star. All candidate companions seen near Vega are visible in this field. The dashed line indicates a radius of 15 arcseconds from the primary star, and the arrow a length of 3".

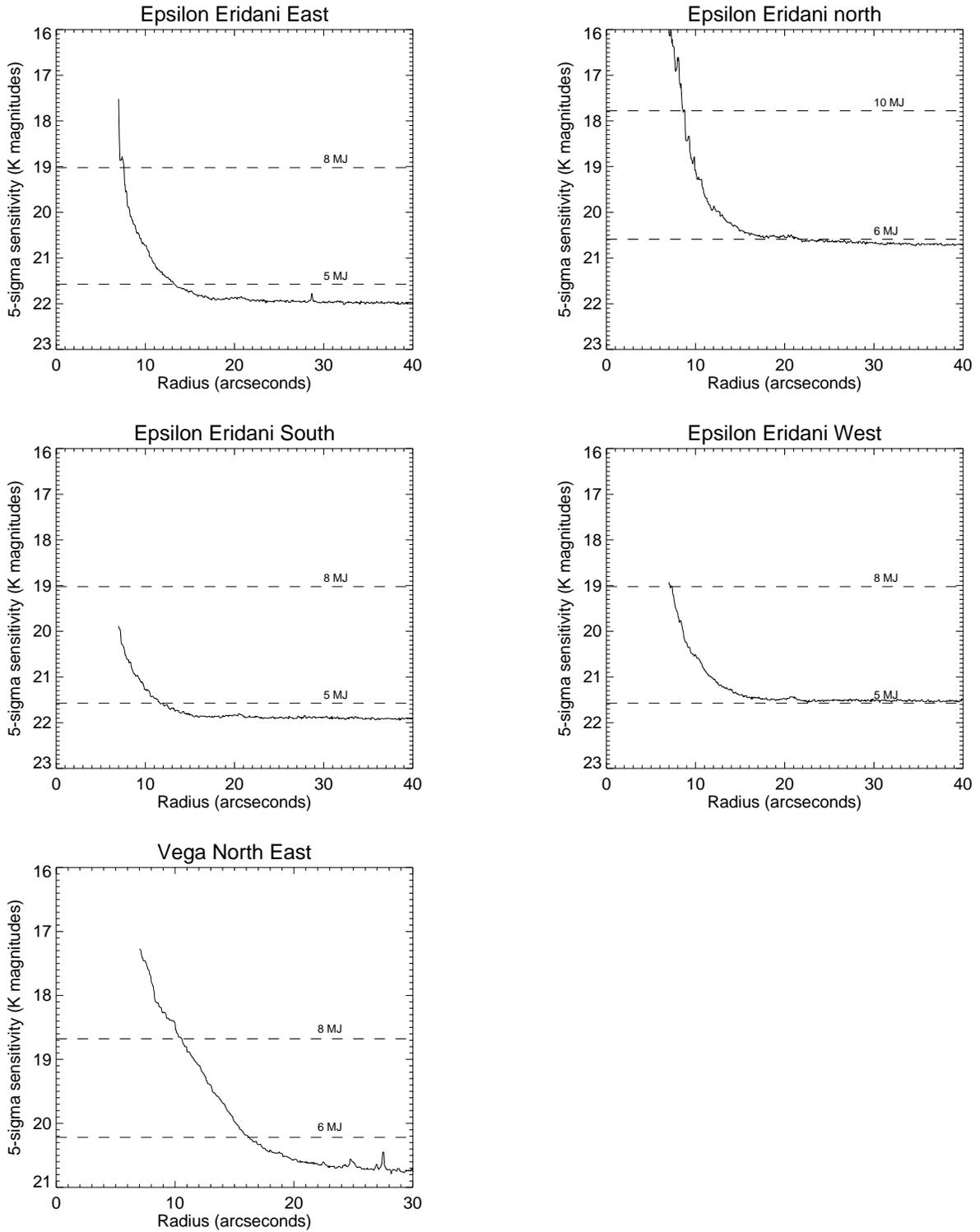

Figure 4: Five-sigma sensitivity of the Epsilon Eridani images (a-d) and Vega images (e). Horizontal lines show the magnitudes of extrasolar planets from the models of Burrows et al (1997).